\documentclass[conference,table,usenames,dvipsnames,svgnames]{IEEEtran}
\pdfoutput=1  

\usepackage[T1]{fontenc}

\usepackage[style=ieee,backend=biber,bibencoding=utf8,texencoding=ascii]{biblatex}
\usepackage{amsmath}
\usepackage{amssymb}
\usepackage{braket}
\usepackage{bm}
\usepackage{cases}  
\usepackage[binary-units]{siunitx}

\usepackage[basic]{complexity}
\newclass{\MAXSAT}{MaxSAT}
\newclass{\MAXCUT}{MaxCUT}

\usepackage{graphicx}
\usepackage[font=footnotesize]{caption}
\usepackage{subcaption}
\usepackage{hyperref}
\usepackage[capitalise]{cleveref} 
\usepackage{csquotes}
\usepackage{booktabs}

\usepackage{float}
\usepackage{xspace}

\DeclareCaptionLabelFormat{nocaption}{}

\title{Finding the optimal Nash equilibrium\\in a discrete Rosenthal congestion game\\using the Quantum Alternating Operator Ansatz}

\author{
	\IEEEauthorblockN{
		Mark Hodson\IEEEauthorrefmark{1},
		Brendan Ruck\IEEEauthorrefmark{1},
		Hugh (Hui Chuan) Ong\IEEEauthorrefmark{2},
		Stefan Dulman\IEEEauthorrefmark{2},
		David Garvin\IEEEauthorrefmark{1}
	}
	\IEEEauthorblockA{
		\IEEEauthorrefmark{1}Rigetti Computing
	}
	\IEEEauthorblockA{
		\IEEEauthorrefmark{2}Commonwealth Bank of Australia
	}
}

\begin{document}


\IEEEtitleabstractindextext{%
	\begin{abstract}
		This paper establishes the tractability of finding the optimal Nash equilibrium, as well as the optimal social solution, to a discrete congestion game using a gate-model quantum computer. The game is of the type originally posited by Rosenthal in the 1970's. To find the optimal Nash equilibrium, we formulate an optimization problem encoding based on potential functions and path selection constraints, and solve it using the Quantum Alternating Operator Ansatz. We compare this formulation to its predecessor, the Quantum Approximate Optimization Algorithm. We implement our solution on an idealized simulator of a gate-model quantum computer, and demonstrate tractability on a small two-player game. This work provides the basis for future endeavors to apply quantum approximate optimization to quantum machine learning problems, such as the efficient training of generative adversarial networks using potential functions.
	\end{abstract}
}

\maketitle

\IEEEdisplaynontitleabstractindextext
\IEEEpeerreviewmaketitle



\section{Introduction}
\label{sec:machine_learning_introduction}

Gate-model Noisy, Intermediate-Scale Quantum (NISQ) \cite{Preskill2018} computers are becoming increasingly available in the cloud, and of sufficient scale and fidelity to run interesting quantum algorithms. Quantum algorithms such as the Quantum Approximate Optimization Algorithm of \cite{Farhi2014} and the Quantum Alternating Operator Ansatz of \cite{Hadfield2019}, collectively QAOA, are able to execute on NISQ hardware and provide the potential of advantage at larger scales. Mapping of industry applications onto quantum algorithms has begun, with particular interest in how hybrid classical-quantum techniques might assist machine learning applications across sectors \cite{PerdomoOrtizEtAl2018}.

In this paper we have brought together machine learning practitioners, financial quantitative analysts and quantum software technologists to investigate how quantum approximate optimization might assist training a generative adversarial network (GAN). Financial services organizations are exploring GANs as a means to generate synthetic data \cite{Kondratyev2019}, allowing machine learning models to be developed without risking real customer data. GANs are also being trained to identify fraud \cite{Zenati2018}, using techniques that may extend to the prediction of network infrastructure issues. Trading \cite{Koshiyama2019} and risk management strategies \cite{Shah2018} are also being learned using GANs, leveraging their expected robustness to changes in market environment.

One challenge facing GAN users is that current gradient-descent methods can fail to converge to the optimal Nash equilibrium in the zero-sum training game. This results in a less accurate model. The challenge is increased in situations where the model includes discrete decision or categorical variables, for which gradients must be synthesized, and where the model presents a highly multi-modal behavior, for which mode collapse becomes an issue \cite{Hui2018}.

In considering how quantum computing might provide a solution to the challenge of training a GAN, we return to discrete neural network models previously considered intractable classically. We lay the groundwork for a new approach by first implementing and experimentally testing methods for calculating the optimal Nash equilibrium and the optimal social solution for a discrete congestion game \cite{Rosenthal1973} using QAOA. We describe the relevance of this application to the broader goals of quantum machine learning, its formulation in both QAOA variants, experimental results, lessons learned, and avenues for further development.


\section{Preliminaries}
\label{sec:preliminaries}

We summarize quantum algorithms and identities upon which this research is based.

\subsection{Binary to spin system identity}

Conversion from a binary system based on $x \in \lbrace 0, 1 \rbrace$ to a spin system based on $s \in \lbrace -1, +1 \rbrace$ is afforded by substitution using the identity

\begin{equation}
	\label{eq:binary_to_spin_system_identity}
	s = 2x-1
\end{equation}

\subsection{Penalty functions for soft constraints}

A real-valued ($\mathbf{x} \in \mathbb{R}^N$, $c \in \mathbb{R}$) equality constraint of the form

\begin{equation*}
	f(\mathbf{x}) = c
\end{equation*}

can be converted to a form that can be solved by unconstrained optimization using a \emph{penalty function} as

\begin{equation}
	\label{eq:penalty_functions_for_soft_constraints}
	P(\mathbf{x}) = A \left( f(\mathbf{x}) - c \right)^2
\end{equation}

where \begin{itemize}
	\item $P(\mathbf{x}) = 0$ when the constraint is met
	\item $P(\mathbf{x}) > 0$ when the constraint is violated
	\item $A \in \mathbb{R} \mid A > 0$ is a penalty scaling coefficient
\end{itemize}

\subsection{Quantum approximate optimization}
\label{sec:preliminaries_quantum_approximate_optimization}

The Quantum Approximate Optimization Algorithm \cite{Farhi2014} has been extended \cite{Wang2017} to minimize a polynomial cost function with real-valued coefficients and discrete solution variables. Discrete solution variables can be defined as a binary system $\mathbf{x} \in \lbrace 0, 1 \rbrace^N$ or spin system $\mathbf{s} \in \lbrace -1, +1 \rbrace^N$ of $N$ variables.

In its canonical form, QAOA's polynomial cost function is a sum-of-products expression with each product term interacting between $0$ and $N$ of the solution variables. If we consider each possible unique interaction then, using the binomial theorem, the maximum number of terms is $2^N$. We observe that the total number of non-zero polynomial coefficients in a tractable problem formulation must scale favorably with respect to the problem size, as each coefficient must be calculated during pre-processing and input as a parameter to the QAOA circuit.

Many interesting optimization problems have at most quadratic terms in the polynomial cost function \cite{Lucas2014}. We restrict ourselves to quadratic problems formulated as a spin system. This results in the Ising model optimization cost function familiar to quantum annealing, as

\begin{equation}
	\label{eq:ising_model_cost_function}
	C(\mathbf{s}) = c + \sum_{i=1}^N h_i s_i + \sum_{i=1}^N \sum_{j=i+1}^N J_{ij} s_i s_j
\end{equation}

where \begin{itemize}
	\item $c \in \mathbb{R}$ is a constant term
	\item $h_i \in \mathbb{R}$ is a coefficient of the \emph{bias vector} $\mathbf{h}$
	\item $J_{ij} \in \mathbb{R}$ is a coefficient of the upper-triangular \emph{coupling matrix} $J$
\end{itemize}

The execution of QAOA is as a \emph{variational algorithm} where circuit parameters $\mathbf{\beta} \in [ 0, \pi ]^p$ and $\mathbf{\gamma} \in [ 0, 2\pi ]^p$ are varied to minimize the expectation value $\langle \psi_1 | C | \psi_1 \rangle$, and so measure \enquote{good} solutions with high probability where

\begin{equation}
	\label{eq:qaoa_initial_state}
	| \psi_0 \rangle = | + \rangle ^ {\otimes N}
\end{equation}

is the initial state of the system, and

\begin{equation}
	\label{eq:qaoa_final_state}
	| \psi_1 \rangle = \left( \prod_{\alpha = p}^1 U(B, \beta_{\alpha}) \, U(C, \gamma_{\alpha}) \right) | \psi_0 \rangle
\end{equation}

is the final state of the system.

In \cref{eq:qaoa_final_state}, unitary evolution occurs via two exponentiated operators: $U(B, \beta_{\alpha}) = e^{-i \beta_{\alpha} B}$ and $U(C, \gamma_{\alpha}) = e^{-i \gamma_{\alpha} C}$, with $i = \sqrt{-1}$ being the imaginary number. The gates applied in a quantum computer iterate as $\alpha = 1, \cdots, p$ due to the right-associativity of these operations. Parameter $p \in \mathbb{N}$ is the number of parameterized repetitions in the resulting quantum circuit, and relates linearly to its depth. The quantum circuit hyper-parameter space of $\beta$ and $\gamma$ also increases linearly with $p$, and is optimized classically.

For our case of the Ising model cost function in \cref{eq:ising_model_cost_function}, the \emph{cost operator} is defined in the Pauli-Z basis ($\sigma^z$) as

\begin{equation}
	\label{eq:qaoa_ising_cost_operator}
	C = \sum_{i=1}^N h_i \sigma_i^z + \sum_{i=1}^N \sum_{j=i+1}^N J_{ij} \sigma_i^z \sigma_j^z
\end{equation}

and is not unitary but is Hermitian, allowing it to be used as the expectation value observable.

For unconstrained optimization problems, the Quantum Approximate Optimization Algorithm defines a \emph{mixing operator} that explores all $2^N$ combinatorial solutions. It is defined in the Pauli-X basis ($\sigma^x$) \cite[Equation (3)]{Farhi2014} in a way physically similar to quantum annealing, as

\begin{equation}
	\label{eq:qaoa_farhi_mixing_operator}
	B = \sum_{i=1}^N \sigma_i^x
\end{equation}

For constrained optimization problems, the Quantum Alternating Operator Ansatz suggests $B$ be designed to constrain the feasible subspace of solutions. One example of this is a \emph{parity mixer}, which uses alternating application of XY mixers to odd and even spin subsets \cite[Equations (7)-(9)]{Hadfield2019}, we summarize as

\begin{equation}
	\label{eq:qaoa_parity_mixing_operator}
	U(B, \beta_{\alpha}) = U(B_{\text{last}}, \beta_{\alpha}) \, U(B_{\text{even}}, \beta_{\alpha}) \, U(B_{\text{odd}}, \beta_{\alpha})
\end{equation}

with

\begin{equation*}
	B_{\text{odd}} = \sum_{a \text{ odd}}^{N-1} \sigma_{a}^x \sigma_{a+1}^x + \sigma_{a}^y \sigma_{a+1}^y
\end{equation*}

\begin{equation*}
	B_{\text{even}} = \sum_{a \text{ even}}^N \sigma_{a}^x \sigma_{a+1}^x + \sigma_{a}^y \sigma_{a+1}^y
\end{equation*}

\begin{equation*}
	B_{\text{last}} = \Big\{ \begin{array}{l}
		\sigma_N^x \sigma_1^x + \sigma_N^y \sigma_1^y, \; N \text{ odd} \\[0.5em]
		I, \; N \text{ even}
	\end{array}
\end{equation*}

and where $I$ is the identity transform and all arithmetic is modulo $N$.

Such a mixer can, for example, be used to realize \emph{one-hot encoding} of categorical variables and has the potential to improve application performance over the original QAOA.

\section{Congestion Game Application}
\label{sec:machine_learning_application}

A congestion game is a class of game-theoretic problem involving players, resources, and a utility function that depends on the number of players sharing each resource -- the congestion. A congestion game is a specialization of a potential game, and it is the use of potential functions that will form the basis of our formulation.

\subsection{Previous work}
\label{sec:congestion_game_application_previous_work}

The congestion game was originally introduced in \cite{Rosenthal1973}, and is a type of game that is guaranteed to possess at least one pure Nash equilibrium. \cite{Meyers2008} showed that finding a Nash equilibrium in an asymmetric network congestion game with linear delay functions is PLS-complete, and finding the social optimum is NP-hard. \cite{YANNAKAKIS200971} showed that finding a Nash equilibrium in a congestion game is PLS-complete in general, even for two players, which is exponential in the worst case to solve. However, such solutions may not be the optimal Nash equilibrium, being the Nash equilibrium with the lowest combined delay for all players. \cite{DelPia2016} proved the global minimum to a symmetric congestion game is the socially optimal Nash equilibrium. \cite{Conitzer:2003:CRN:1630659.1630770, Conitzer:2006:COS:1134707.1134717} showed that finding a Nash equilibrium with the maximum utility for even a single player in a two-player game is NP-hard. \cite{Dutting2015} confirmed some of these complexity results, and provided some small-scale network games that were useful in early development.

Our choice to focus on the use of potential functions is informed by recent developments \cite{OliehoekEtAl2018}, in which a method for training a GAN is proposed that yields a single optimal Nash equilibrium by using potential functions. Another method in \cite{OliehoekEtAl2018} considers training a GAN explicitly as a game with mixed strategies, and is shown to avoid mode collapse. \cite{GoodfellowEtAl2014} is the seminal reference for generative adversarial networks, and identifies training a GAN to be a zero-sum game that results in a Nash equilibrium. \cite{MondererShapley1996} extended \cite{Rosenthal1973} to show that congestion games and potential games are equivalent.

Quantum adversarial approaches to machine learning have started to be developed by the quantum computing research community. \cite{PerdomoOrtizEtAl2018} identifies finding the Nash equilibrium of a game as one of a class of industry-relevant applications that could benefit from quantum-assisted machine learning (QAML). \cite{LloydWeedbrook2018} introduces quantum generative adversarial networks (QuGANs) and concludes that quantum adversarial networks may exhibit an exponential advantage over classical adversarial networks, when the data, the generator and the discriminator are all quantum. \cite{DallaireDemersKilloran2018} constructs a GAN using quantum circuits, and shows a technique for computing gradients used during learning. \cite{BenedettiEtAl2018} derives an adversarial algorithm to approximate quantum pure states, and uses resilient back-propagation to overcome the small observed gradients to improve the optimization of generator and discriminator networks.

\subsection{Contribution}
\label{sec:congestion_game_application_contribution}

Our contribution in this work is the experimental evaluation of the tractability of finding the optimal Nash equilibrium in a discrete asymmetric-network congestion game. We implement this game using QAOA on a simulator of a gate-model quantum computer. We design and implement a soft-constraint formulation based on \cite{Farhi2014}, and compare it to a hard-constraint formulation based on \cite{Hadfield2019}. We choose an idealized simulator of a gate-model quantum computer \cite{Smith2017} as a first step in understanding algorithm and application performance, and with the intent of evaluating on NISQ computers in the future \cite{Mahabubul2019}. To our knowledge, no work using QAOA or other QAML techniques to calculate the optimal Nash equilibrium in a game has been published.

\subsection{Relevance}
\label{sec:congestion_game_relevance}

Generative adversarial networks can be trained and used for purposes including classification, where the discriminator network is the product of interest, or for synthetic data generation, where the generator network is the product of interest. Applications for GAN-based machine learning in financial services include synthetic data generation \cite{Kondratyev2019}, market risk management \cite{Shah2018}, risk factor analysis \cite{Hadad2017}, trading strategies \cite{Koshiyama2019}, and detecting fraud and other anomalies \cite{Zenati2018}. Numerous applications of GAN-based machine learning exist in other sectors.

The optimal Nash equilibrium is an important concept in training a GAN. A GAN involves two neural networks competing against each other. The discriminator network tries to accurately discriminate features in observed data. The generator network tries to generate statistically indistinguishable data to fool the discriminator. In this competition the training outcome is hampered if the two networks settle into a sub-optimal Nash equilibrium. This reduces the accuracy of the resulting discriminator, and of the synthetic data produced by the generator.

The Nash equilibrium also finds utility in game-theoretic modeling of human behaviors in traffic and other resource management activities where players act in their own self-interest, but without knowledge of the others' strategies. Understanding what is the optimal solution if players work cooperatively, compared to the optimal competitive Nash equilibrium, provides an indication of how well designed the rules of the game and topology of the network are to achieve a socially desirable outcome.


\section{Congestion Game Formulation}
\label{sec:machine_learning_formulation}

The canonical definition of a congestion game is as a tuple $(\mathcal{N}, \mathcal{R}, \mathcal{S}, d)$, where

\begin{itemize}
	\item $\mathcal{N} = \lbrace 1, \cdots, n \rbrace$ is the set of $n$ players
	\item $\mathcal{R} = \lbrace 1, \cdots, r \rbrace$ is the set of $r$ resources
	\item $\mathcal{S} = \mathcal{S}_1 \times \cdots \times \mathcal{S}_n$ is the strategy space of the game, where $\mathcal{S}_i \subseteq 2^{\mathcal{R}} \setminus \varnothing$ is the strategy space\footnote{The strategy space uses the power set notation $2^{\mathcal{R}}$, which describes every possible combination of choosing, or not choosing, to use a resource.} for player $i \in \mathcal{N}$
	\item $d = (d_1, \cdots, d_r)$ models the congestion, where $d_k: \mathbb{N} \mapsto \mathbb{R}$ is the delay function for using resource $k \in \mathcal{R}$
\end{itemize}

For a specific application, this definition must be specialized to the subset of resources available to each player. For a traffic congestion game each player is given an origin and destination, and is limited in their path by the topology of the network. In this paper we approach a two-player asymmetric network congestion game. This is representative of two players traveling on a road network with different origins or destination locations, and can be NP-hard to calculate both the optimal Nash equilibrium \cite{Conitzer:2003:CRN:1630659.1630770}, and the optimal social solution \cite{Meyers2008}. Excluded from this paper is a theoretical treatment of the computational complexity of this specific case. However we do analyze the limiting behavior of the QAOA cost functions as an initial indicator of tractability at scale, the details of which are presented in \cref{sec:machine_learning_appendix_big_o}.

A solution to a congestion game is a set of actions selected by all players

\begin{equation}
	\label{eq:congestion_game_action_space}
	\mathcal{A} = \lbrace \mathcal{A}_1, \cdots, \mathcal{A}_n \rbrace
\end{equation}

Each player's action can be described as the set of resources utilized in the network

\begin{equation}
	\label{eq:congestion_game_action_player}
	\mathcal{A}_i \subseteq \mathcal{R}, \; i \in \mathcal{N}
\end{equation}

and results in a utility for that player that is the combined delay across the resources utilized\footnote{In the model of a traffic congestion game a player's utility is their total travel time, and the objective of the game is to minimize this.}

\begin{equation}
	\label{eq:congestion_game_player_utility_A}
	u_i(\mathcal{A}) = \sum_{k \in \mathcal{A}_i} d_k(n_k(\mathcal{A}))
\end{equation}

where the delay for each resource depends on the number of players utilizing it

\begin{equation}
	\label{eq:congestion_game_number_of_players_A}
	n_k(\mathcal{A}) = | \lbrace i \in \mathcal{N} \mid k \in \mathcal{A}_i \rbrace |
\end{equation}

The combined utility for all players is then

\begin{equation}
	\label{eq:congestion_game_combined_utility_A}
	u(\mathcal{A}) = \sum_{i \in \mathcal{N}} u_i(\mathcal{A})
\end{equation}

From this we can define the optimal social solution as

\begin{equation*}
	\mathcal{A}_{\text{social}} = \underset{\mathcal{A} \in \mathcal{S}}{\text{argmin}} \; u(\mathcal{A})
\end{equation*}

which minimizes the combined delay for all players, and can be written more explicitly as

\begin{equation}
	\label{eq:congestion_game_A_social}
	\mathcal{A}_{\text{social}} = \underset{\mathcal{A} \in \mathcal{S}}{\text{argmin}} \; \sum_{k \in \mathcal{R}} n_k(\mathcal{A}) d_k(n_k(\mathcal{A}))
\end{equation}

Using the potential function approach of \cite[Eq. (3.2)]{MondererShapley1996} we can define the optimal Nash equilibrium as

\begin{equation}
	\label{eq:congestion_game_A_Nash}
	\mathcal{A}_{\text{Nash}} = \underset{\mathcal{A} \in \mathcal{S}}{\text{argmin}} \; \sum_{k \in \mathcal{R}} \sum_{j}^{n_k(\mathcal{A})} d_k(j)
\end{equation}

which is the socially optimal Nash equilibrium since any change to the solution incurs a change in optimization value that is equal to the change in utility for each affected player. This is confirmed by \cite{DelPia2016} where the same formulation as \cref{eq:congestion_game_A_Nash} was used as part of analysis of symmetric congestion games.

\subsection{Path model of player strategy space}

In network congestion games the strategy space for a player to utilize resources is constrained by the network topology. In a traffic congestion game the set of paths that a player might utilize to travel from source to destination can be considered to be small and fixed, based on that player's own knowledge of the road network, augmented by consumer navigation applications. In designing the model that represents a player's use of network resources, the na\"{i}ve approach of an independent binary decision variable for each player and each resource is inefficient because it requires $\mathcal{O}(nr)$ variables with a very large number of constraints. We therefore assume a path-finding algorithm exists that can provide a small and fixed number of available paths for each player, and we design the player's strategy space based on these alternatives.

We define the existing path-finding function $f_{\text{path}}$ as

\begin{equation*}
	f_{\text{path}} : \lbrace \mathcal{N}, 2^{\mathcal{R}} \rbrace \rightarrow \lbrace \text{True}, \text{False} \rbrace
\end{equation*}

and then define the available player strategies as exactly the subset that were suggested by the path-finding function as

\begin{equation}
	\label{eq:congestion_game_player_strategies}
	\mathcal{S}_i = \lbrace \mathcal{A}_i \in 2^{\mathcal{R}} \mid f_{\text{path}}(i, \mathcal{A}_i) \rbrace, \; i \in \mathcal{N}
\end{equation}

\subsection{Linear model of resource congestion}

In network congestion games each resource delay function $d_k$ is an arbitrary non-decreasing function that maps the number of players sharing a resource onto the delay that each player incurs in using the resource. However, arbitrary functions can be expensive to represent using polynomial cost functions available to the quantum approximate optimization techniques of \cref{sec:preliminaries_quantum_approximate_optimization}. We therefore restrict our formulation to linear resource delay functions, which does not reduce the complexity of finding a Nash equilibrium in the analysis of \cite{Meyers2008}.

We define each resource delay function $d_k$ as a linear function

\begin{equation}
    \label{eq:congestion_game_resource_delay}
    d_k(x) = a_k + b_k x, \; x \in \mathbb{N}, \; k \in \mathcal{R}
\end{equation}

where $a_k \in \mathbb{R}$ and $b_k \in \mathbb{R}$ are fixed resource delay coefficients, and $x$ is the number of players utilizing the resource.

\subsection{Player strategy space encoding}

We encode each player's strategy space from \cref{eq:congestion_game_player_strategies} using binary decision variables $z_{i,j}$ that denote whether player $i$ chooses path $j$, and build the binary solution vector

\begin{equation}
	\label{eq:congestion_game_solution_variable_z}
	\mathbf{z} = \lbrace z_{i,j} \mid i \in \mathcal{N}, j \in \mathcal{S}_i \rbrace, \; z_{i,j} \in \lbrace 0, 1 \rbrace
\end{equation}

which using \cref{eq:binary_to_spin_system_identity} can be converted to the spin-system solution vector

\begin{equation}
	\label{eq:congestion_game_solution_variable_s}
	\mathbf{s} = \lbrace s_{i,j} \mid i \in \mathcal{N}, j \in \mathcal{S}_i \rbrace, \; s_{i,j} \in \lbrace -1, +1 \rbrace
\end{equation}

Each player may only choose a single path. This is enforced using linear constraints

\begin{equation}
	\label{eq:congestion_game_path_constraint_z}
	\sum_{j \in \mathcal{S}_i} z_{i,j} = 1, \; \forall i \in \mathcal{N}
\end{equation}

or equivalently by \cref{eq:binary_to_spin_system_identity} as

\begin{equation}
	\label{eq:congestion_game_path_constraint_s}
	\sum_{j \in \mathcal{S}_i} s_{i,j} = 2 - |\mathcal{S}_i|, \; \forall i \in \mathcal{N}
\end{equation}

By meeting this constraint, each player's action is defined as the single strategy selected by the solution

\begin{equation}
	\label{eq:congestion_game_player_actions}
	\mathcal{A}_i = j \in \mathcal{S}_i \ni z_{i,j} = 1, \; i \in \mathcal{N}
\end{equation}

and so the solution to the game $\mathcal{A}$ is encoded in $\mathbf{z}$.

\subsection{Player utility function encoding}

We can calculate the number of players using a resource $k \in \mathcal{R}$ from \cref{eq:congestion_game_number_of_players_A} by summing the players whose chosen paths in \cref{eq:congestion_game_solution_variable_z} include that resource

\begin{equation}
	\label{eq:congestion_game_number_of_players_z}
	n_k(\mathbf{z}) = \sum_{i \in \mathcal{N}} \sum_{j \in \mathcal{S}_i \mid k \in j} z_{i,j}
\end{equation}

or, equivalently by \cref{eq:binary_to_spin_system_identity} as

\begin{equation}
	\label{eq:congestion_game_number_of_players_s}
	n_k(\mathbf{s}) = \frac{1}{2} \sum_{i \in \mathcal{N}} \sum_{j \in \mathcal{S}_i \mid k \in j} (1 + s_{i,j})
\end{equation}

From this, we can now calculate the delay for a resource $k \in \mathcal{R}$ by substitution into \cref{eq:congestion_game_resource_delay}, as

\begin{equation}
    \label{eq:congestion_game_resource_delay_s}
    d_k(n_k(\mathbf{s})) = a_k + \frac{b_k}{2} \left( \sum_{i \in \mathcal{N}} \sum_{j \in \mathcal{S}_i \mid k \in j} (1 + s_{i,j}) \right)
\end{equation}

The utility for a player who has selected their action as in \cref{eq:congestion_game_player_actions} can now be calculated from \cref{eq:congestion_game_player_utility_A} as

\begin{equation}
    \label{eq:congestion_game_player_utility_s}
    u_i(\mathbf{s}) = \sum_{k \in \mathcal{A}_i} \left( a_k + \frac{b_k}{2} \left( \sum_{i \in \mathcal{N}} \sum_{j \in \mathcal{S}_i \mid k \in j} (1 + s_{i,j}) \right) \right)
\end{equation}

\subsection{Path constraint (soft constraint form)}

The path selection constraint of \cref{eq:congestion_game_path_constraint_s} can be converted to a penalty function using \cref{eq:penalty_functions_for_soft_constraints} as

\begin{equation}
	\label{eq:congestion_game_C_path_s}
	C_{\text{path}}(\mathbf{s}) = A \sum_{i \in \mathcal{N}} \left( \sum_{j \in S_i} s_{i,j} + |\mathcal{S}_i| - 2 \right)^2
\end{equation}

which is a summation of the $n$ independent player path constraints, where the penalty scaling coefficient $A \in \mathbb{R}$ is to be determined experimentally.

This penalty function contains at most $\mathcal{O}(n p^2)$ quadratic terms, where $p$ denotes the number of paths available to each player. Refer to the analysis in \cref{sec:analysis_of_the_soft_constraint_penalty_function} for details.

\subsection{Path constraint (hard constraint form)}
\label{sec:path_constraint_hard_constraint_form}

\begin{figure}
	\centering
	\includegraphics[width=\linewidth]{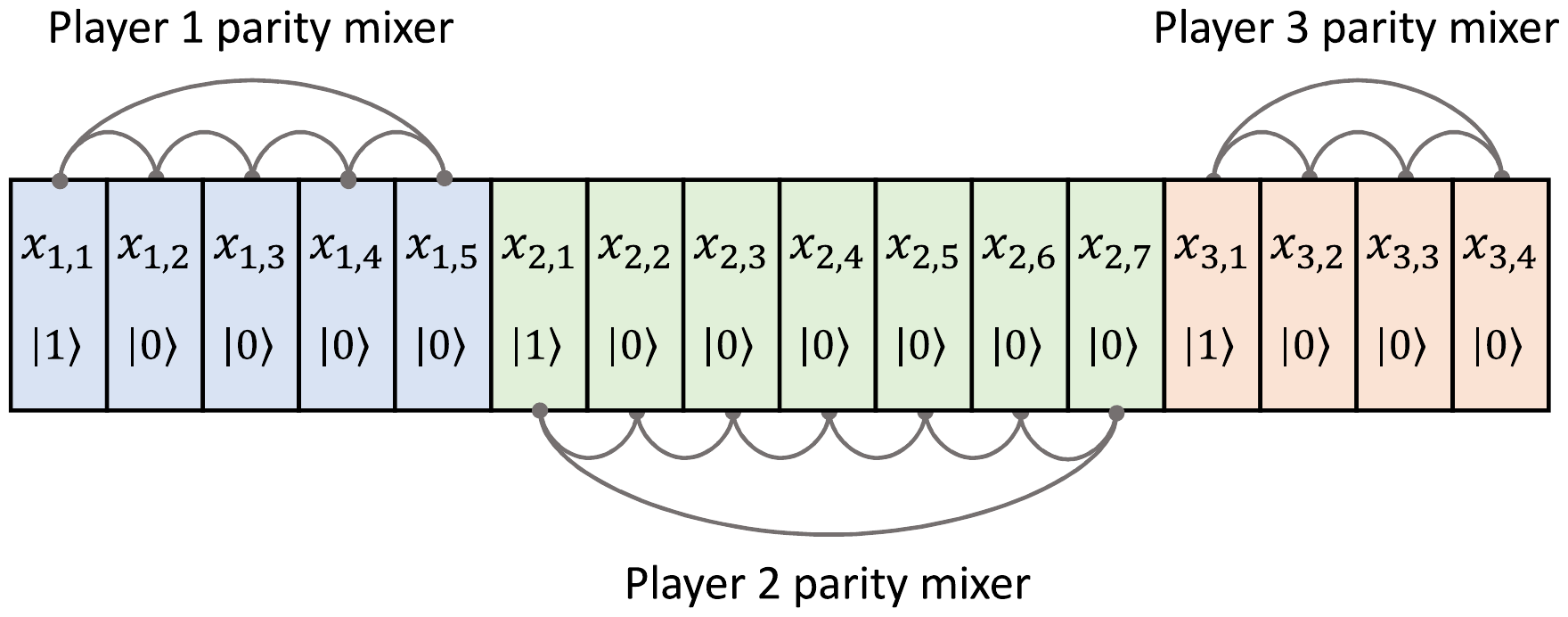}
	\caption{Initial configuration for the hard constraint form of the path selection constraint in a three-player game. The three players have 5, 7 and 4 options, respectively.}
	\label{fig:congestion_game_parity_mixer}
\end{figure}

The path selection constraint of \cref{eq:congestion_game_path_constraint_s} can be realized using $n$ parity ring mixers \cite{Hadfield2019} acting independently on each player's path selection decision variables. Each parity mixer is configured to preserve a Hamming weight of 1 by initialization of the state space into the solution where each player selects the first path option. An illustration of this configuration for a three-player game is provided in \cref{fig:congestion_game_parity_mixer}.

\subsection{Finding the optimal social solution}
\label{sec:finding_the_optimal_social_solution}

The objective of finding the optimal social solution in \cref{eq:congestion_game_A_social} can now be written as a minimization objective over the solution vector $\mathbf{s}$, as

\begin{equation}
	\label{eq:congestion_game_opt_s_social}
	\mathbf{s}_{\text{social}} = \underset{\mathbf{s}}{\text{argmin}} \; C_{\text{social}}(\mathbf{s})
\end{equation}

where

\begin{equation}
	\label{eq:congestion_game_C_s_social}
	C_{\text{social}}(\mathbf{s}) = \sum_{k \in \mathcal{R}} n_k(\mathbf{s}) \, d_k(n_k(\mathbf{s}))
\end{equation}

and where the components of the optimization are defined by \cref{eq:congestion_game_number_of_players_s} and \cref{eq:congestion_game_resource_delay_s}.

This cost function contains at most quadratic terms and can be expanded for a specific network topology. The number of quadratic terms is $\mathcal{O}(r n^2 p^2)$ in the worst case. Refer to the analysis in \cref{sec:analysis_of_the_optimal_social_solution} for details.

\subsection{Finding the optimal Nash equilibrium}
\label{sec:finding_the_optimal_nash_equilibrium}

The objective of finding the optimal Nash equilibrium in \cref{eq:congestion_game_A_Nash} can also be written as a minimization objective over the solution vector $\mathbf{s}$, as

\begin{equation}
	\label{eq:congestion_game_opt_s_Nash}
	\mathbf{s}_{\text{Nash}} = \underset{\mathbf{s}}{\text{argmin}} \; C_{\text{Nash}}(\mathbf{s})
\end{equation}

where

\begin{equation}
	\label{eq:congestion_game_C_s_Nash_early}
	C_{\text{Nash}}(\mathbf{s}) = \sum_{k \in \mathcal{R}} \sum_{j}^{n_k(\mathbf{s})} d_k(j)
\end{equation}

and where the components of the optimization are defined by \cref{eq:congestion_game_number_of_players_s} and \cref{eq:congestion_game_resource_delay}.

This expression contains an inner sum whose number of terms depends on the solution. This form is not programmable on a quantum computer, where it must be possible to calculate the coefficients of the cost function polynomial of \cref{eq:qaoa_ising_cost_operator} and build the quantum circuit. However, the summation over $n_k(\mathbf{s})$ is shown in \cref{sec:analysis_of_the_optimal_nash_equilibrium} to reduce to a triangular number

\begin{equation*}
	T_n = \sum_{k=1}^n k = 1 + 2 + 3 + \cdots + n = \frac{n (n + 1)}{2}
\end{equation*}

and generates the form

\begin{equation}
	\label{eq:congestion_game_C_s_Nash}
	C_{\text{Nash}}(\mathbf{s}) = \sum_{k \in \mathcal{R}} \left( \left( a_k + \frac{b_k}{2} \right) n_k(\mathbf{s}) + \frac{b_k}{2} n_k(\mathbf{s})^2 \right)
\end{equation}

This cost function contains at most quadratic terms and can now be expanded for a specific network topology. The number of quadratic terms is $\mathcal{O}(r n^2 p^2)$ in the worst case. Refer to the analysis in \cref{sec:analysis_of_the_optimal_nash_equilibrium} for details.

\subsection{Soft constraint formulation using the Quantum Approximate Optimization Algorithm}
\label{sec:machine_learning_formulation_qaoa_realization_soft_constraints}

The congestion game application can be realized using an unconstrained optimization approach by combining either the optimal social cost function (\cref{eq:congestion_game_C_s_social}) or the optimal Nash equilibrium cost function (\cref{eq:congestion_game_C_s_Nash}) with the path selection penalty function (\cref{eq:congestion_game_C_path_s}), as

\begin{equation}
	\label{eq:C_social_soft_s}
	C_{\text{social}}^{\text{soft}}(\mathbf{s}) = C_{\text{social}}(\mathbf{s}) + C_{\text{path}}(\mathbf{s})
\end{equation}

and

\begin{equation}
	\label{eq:C_Nash_soft_s}
	C_{\text{Nash}}^{\text{soft}}(\mathbf{s}) = C_{\text{Nash}}(\mathbf{s}) + C_{\text{path}}(\mathbf{s})
\end{equation}

Solving this formulation using the Quantum Approximate Optimization Algorithm involves its execution as described in \cref{sec:preliminaries_quantum_approximate_optimization}. The QAOA cost operator is of the \cref{eq:qaoa_ising_cost_operator} form, generated by substitution of the Pauli-Z operator $\sigma_i^z$ for each spin $s_i \in \mathbf{s}$ in \cref{eq:C_social_soft_s} and \cref{eq:C_Nash_soft_s}. The QAOA unconstrained mixing operator is of the standard \cref{eq:qaoa_farhi_mixing_operator} form.

\subsection{Hard constraint formulation using the Quantum Alternating Operator Ansatz}
\label{sec:machine_learning_formulation_qaoa_realization_hard_constraints}

The congestion game application can also be realized using a constrained optimization approach by using only the optimal social cost function (\cref{eq:congestion_game_C_s_social}) or the optimal Nash equilibrium cost function (\cref{eq:congestion_game_C_s_Nash}), as

\begin{equation}
	\label{eq:C_social_hard_s}
	C_{\text{social}}^{\text{hard}}(\mathbf{s}) = C_{\text{social}}(\mathbf{s})
\end{equation}

and

\begin{equation}
	\label{eq:C_Nash_hard_s}
	C_{\text{Nash}}^{\text{hard}}(\mathbf{s}) = C_{\text{Nash}}(\mathbf{s})
\end{equation}

and whose path selection constraint is realized as described in \cref{sec:path_constraint_hard_constraint_form}.

Solving this formulation using the Quantum Alternating Operator Ansatz involves its execution as described in \cref{sec:preliminaries_quantum_approximate_optimization}. The QAOA cost operator remains of the \cref{eq:qaoa_ising_cost_operator} form, generated by substitution of the Pauli-Z operator $\sigma_i^z$ for each spin $s_i \in \mathbf{s}$ in \cref{eq:C_social_hard_s} and \cref{eq:C_Nash_hard_s}. The $n$ constrained mixing operators are of the \cref{eq:qaoa_parity_mixing_operator} form, with each mixer $i \in \mathcal{N}$ applying to the set of spin variables $\lbrace s_{i,j} \mid j \in \mathcal{S}_i \rbrace$, and the initial feasible state being $|1\rangle$.


\section{Congestion Game Experimental Results}
\label{sec:machine_learning_experimentation}

We execute the congestion game formulation upon an idealized simulator of a gate-model quantum computer to assess its tractability prior to execution on NISQ hardware \cite{Smith2017}. We do this in a sequence of steps designed to verify the application is implemented correctly.

\subsection{Experimental data}

\begin{figure}
	\centering
	\includegraphics[scale=0.03]{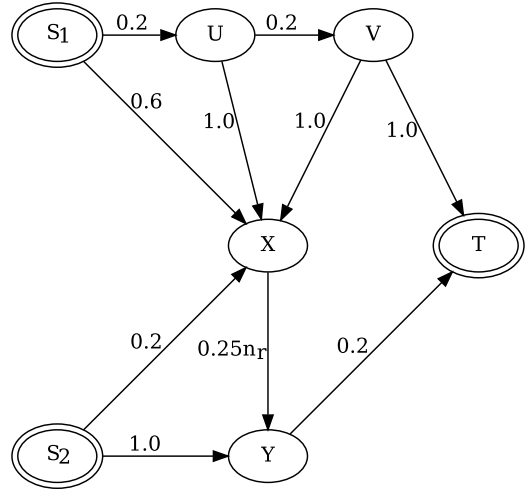}
	\caption{The asymmetric network game used for the experiments. In this game, player A starts at node $S_1$ and travels to node $T$, while player B starts at node $S_2$ and travels to node $T$. The edge from node $X$ to $Y$ has a delay that is a function of the number of players using the resource, $n_r$.}
	\label{fig:network_game}
\end{figure}

The input data for this experiment is a small asymmetric network congestion game with two players and seven nodes, illustrated in \cref{fig:network_game}. This game was designed such that the combined utility of all players for the optimal social solution was different than that of the optimal Nash equilibrium.

For this game, the path-finding function was deemed to return all possible paths that honor the directed edges in the graph. By inspection, it can be deduced that player A has four available paths, while player B has two.

\subsection{Calculation of penalty scaling}

A simple method to calculate penalty scaling coefficient $A$ in \cref{eq:congestion_game_C_path_s} was adopted, as

\begin{equation}
	\label{eq:A_penalty_max_min}
    A > \text{max}\left[ C(\mathbf{s}) \right] - \text{min}\left[ C(\mathbf{s}) \right]
\end{equation}

where $C(\mathbf{s}) = C_{\text{social}}(\mathbf{s})$ is the unconstrained part of \cref{eq:C_social_soft_s} for finding the optimal social solution, and $C(\mathbf{s}) = C_{\text{Nash}}(\mathbf{s})$ is the unconstrained part of \cref{eq:C_Nash_soft_s} for finding the optimal Nash equilibrium. This ensures that the value of $C(\mathbf{s})$ for any infeasible solution is greater than the energy for all feasible solutions. The performance of this setting for realizing feasible solutions will be validated experimentally.

\subsection{Investigation of the solution space}

We implemented the optimal social solution cost function $C_{\text{social}}(\mathbf{s})$ and optimal Nash equilibrium cost function $C_{\text{Nash}}(\mathbf{s})$ and evaluated them by brute force. Following the formulation of \cref{sec:machine_learning_formulation}, a system involving only 6 spin variables is realized, making this type of analysis possible. This generates a solution space of 64 binary solution vectors, of which only 8 meet the path selection constraint of \cref{eq:congestion_game_path_constraint_s}. The 8 feasible solutions represent the product of independent decisions of player A choosing 1 from 4 paths, and player B choosing 1 from 2 paths.

\begin{figure}
	\centering
	\includegraphics[scale=0.03]{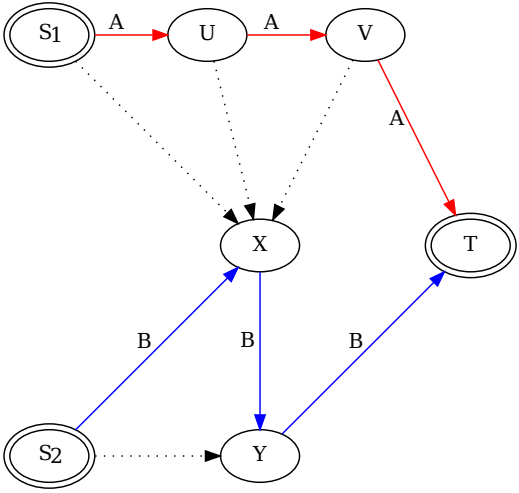}
	\caption{Brute force optimal social solution for the asymmetric network game.}
	\label{fig:network_game_optimum_social}
\end{figure}

\begin{figure}
	\centering
	\includegraphics[scale=0.03]{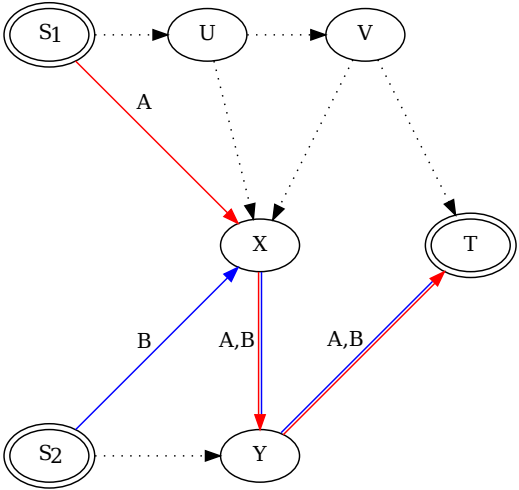}
	\caption{Brute force optimal Nash equilibrium for the asymmetric network game.}
	\label{fig:network_game_optimum_nash}
\end{figure}

\cref{fig:network_game_optimum_social} depicts the optimal social solution, while \cref{fig:network_game_optimum_nash} depicts the optimal Nash equilibrium calculated by brute force. The optimal social solution had a combined utility of 2.05, and involved no common paths. The optimal social solution is not a Nash equilibrium because player A, who has a delay of 1.4 ($S_1 - U - V - T$), has the option of choosing an alternate route that would incur a smaller delay of 1.3 ($S_1 - X - Y - T$). However, if player A were to do this, then player B's delay would increase by more than the benefit to player A. This change results in the optimal Nash equilibrium, which has a combined utility of 2.2.

\subsection{Investigation of a single QAOA circuit}

\begin{figure}
	\centering
	\includegraphics[width=\linewidth]{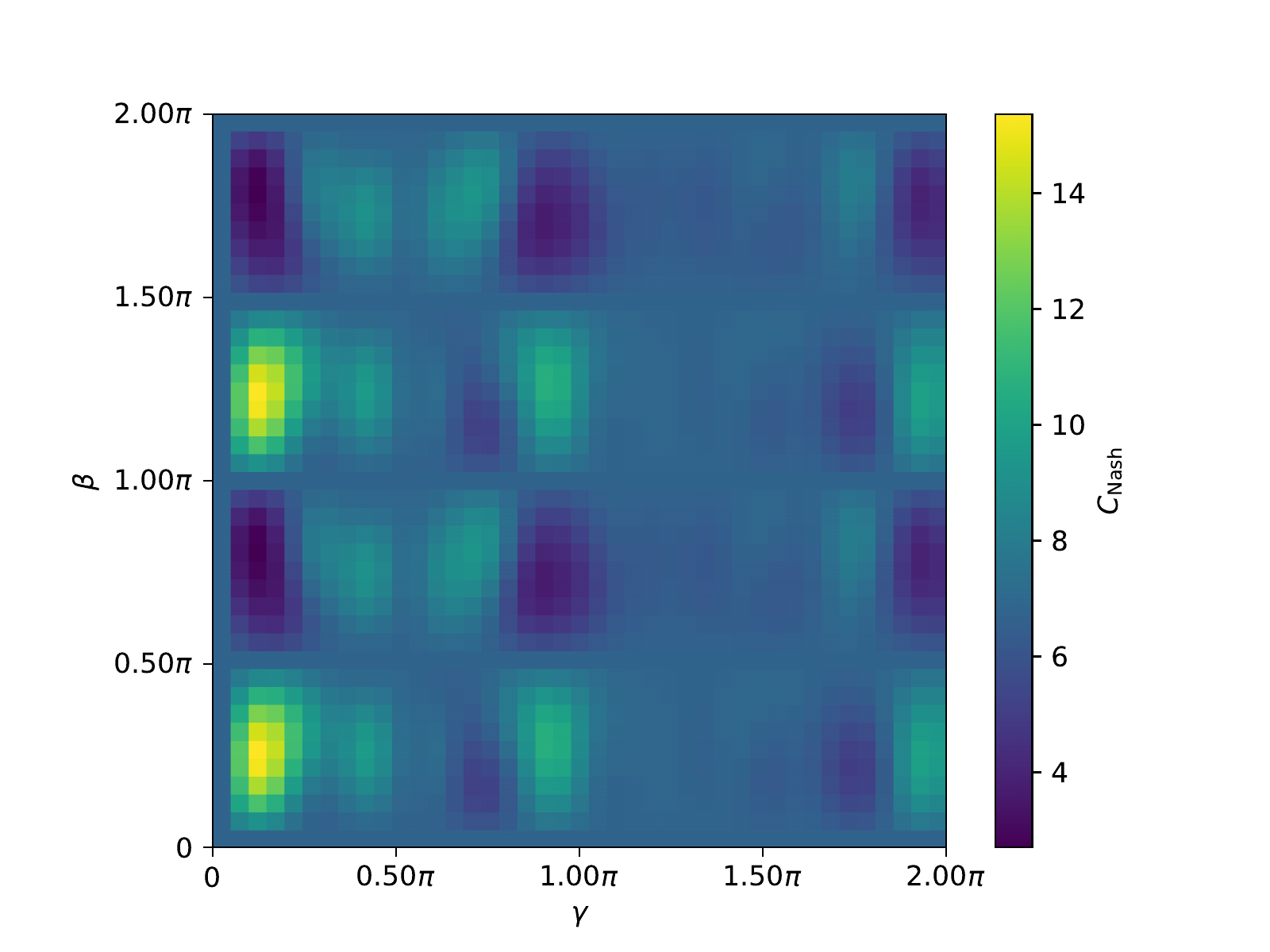}
	\caption{$\langle \psi_1 | C_{\text{Nash}}(\mathbf{s}) | \psi_1 \rangle$ for the asymmetric network game, using the Quantum Approximate Optimization Algorithm with $A = 10$, $p = 1$.}
	\label{fig:gamma_beta_classic_a10}
\end{figure}

Before enabling the outer classical optimization loop of QAOA, we first investigated the behavior for a single iteration of the Quantum Approximate Optimization Algorithm circuit on formulation \cref{eq:C_Nash_soft_s} for the optimal Nash equilibrium. The penalty scaling coefficient is calculated using \cref{eq:A_penalty_max_min} as $A = 10$. We vary $\beta$ and $\gamma$ angles for a $p = 1$ execution of \cref{eq:qaoa_final_state}, and obtain the results in \cref{fig:gamma_beta_classic_a10}. This figure shows good mixing in the quantum state space, indicating the scale of the cost function data is acceptable.

\begin{figure}
	\centering
	\includegraphics[width=\linewidth]{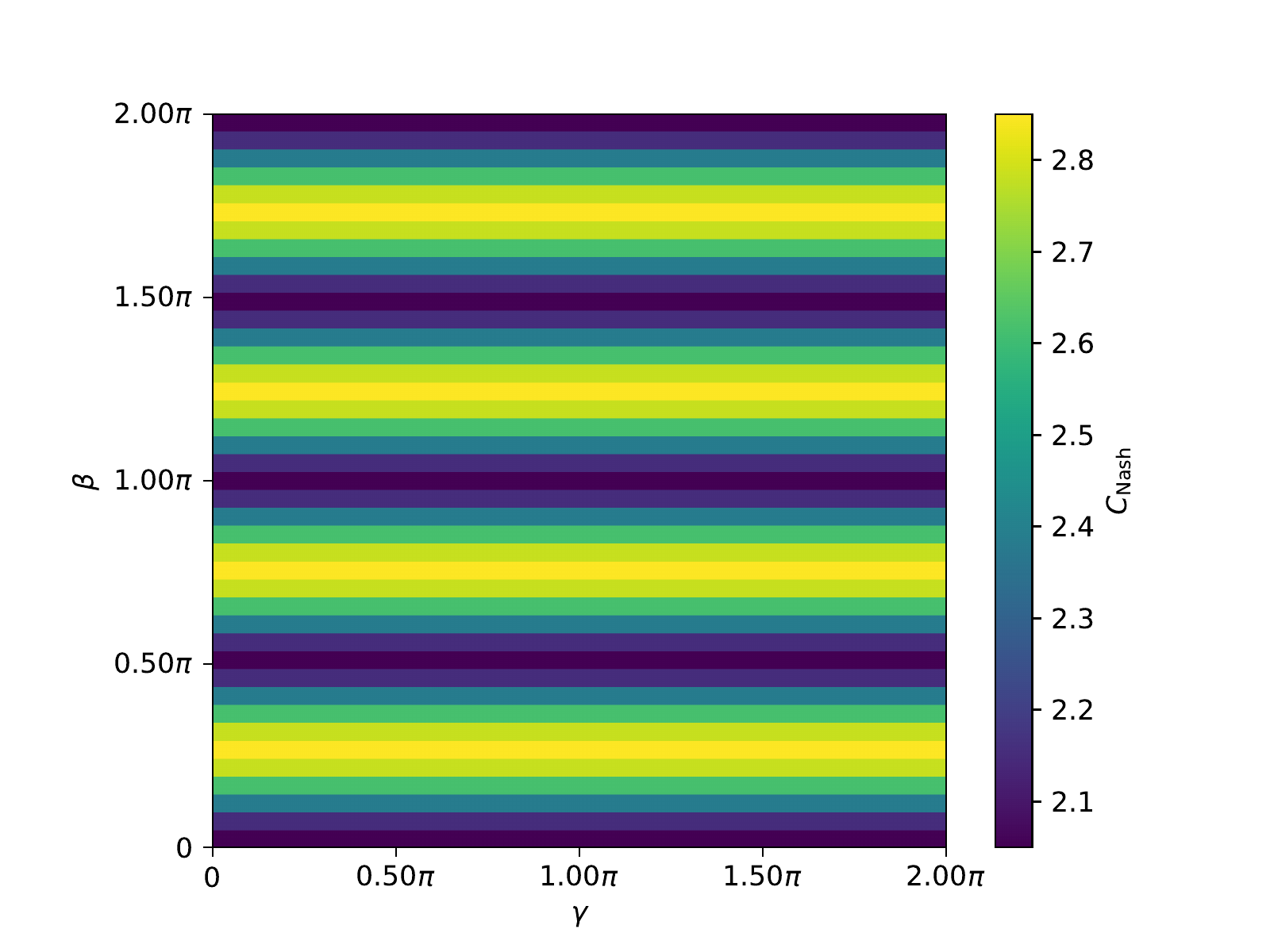}
	\caption{$\langle \psi_1 | C_{\text{Nash}}(\mathbf{s}) | \psi_1 \rangle$ for the asymmetric network game, using the Quantum Alternating Operator Ansatz with $p = 1$ and no initial mixing.}
	\label{fig:nash_hc_single_no_mixed}
\end{figure}

We also investigated the behavior for a single iteration of the Quantum Alternating Operator Ansatz circuit on formulation \cref{eq:C_Nash_hard_s} for the optimal Nash equilibrium. As designed in \cref{fig:congestion_game_parity_mixer} the qubit registers holding the decision spaces for player A and player B were initialized into the $|1\rangle$ state, and in accordance with the normal flow of the QAOA algorithm applied the cost function gates before the mixing operator gates. The $\gamma$ vs. $\beta$ heat-map that resulted is shown in \cref{fig:nash_hc_single_no_mixed}, and has no variation across the $\gamma$ axis. This is consistent with the fact that the $z$-rotations caused by the cost operator gates have no effect on the quantum state when in a computational basis state.

\begin{figure}
	\centering
	\includegraphics[width=\linewidth]{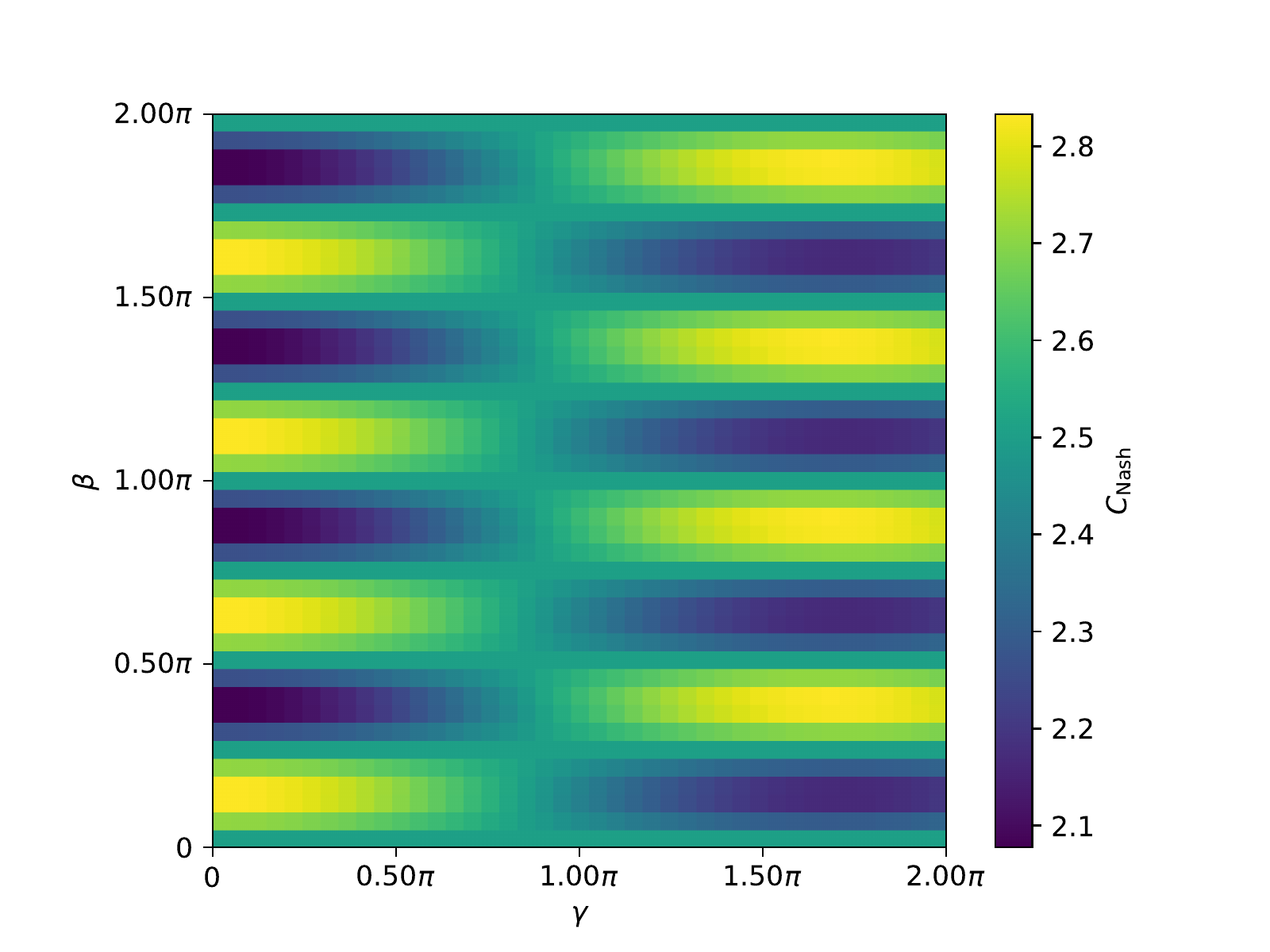}
	\caption{$\langle \psi_1^+ | C_{\text{Nash}}(\mathbf{s}) | \psi_1^+ \rangle$ for the asymmetric network game, using the Quantum Alternating Operator Ansatz with $p = 1$ and an initially mixed state $|\psi_1^+\rangle = U(B, \beta_0) |\psi_1\rangle$ generated with $\beta_0 = \tfrac{\pi}{8}$.}
	\label{fig:nash_hc_single_mixed}
\end{figure}

If an initial mixer $U(B, \beta_0)$ is applied immediately following the register initialization, we create a superposition of feasible states and can then see variation on both $\gamma$ and $\beta$ axes as in \cref{fig:nash_hc_single_mixed}. For the initial mixing, $\beta_0$ was selected as $\tfrac{\pi}{8}$.

\subsection{Evaluation of QAOA in solving the game}

We enable the outer classical optimization loop for both the Quantum Approximate Optimization Algorithm and Quantum Alternating Operator Ansatz formulations of the problem. We perform 10 simulated executions of each using randomly seeded $\beta$ and $\gamma$ angles, and repeat for $p \in \lbrace 1, 3, 5, 8 \rbrace$ parameterized repetitions. We include brute force baseline statistics, generated from a uniform distribution across all possible, and all feasible solutions. In reporting results for these experiments, we refer to the Quantum Approximate Optimization Algorithm solutions to \cref{eq:C_social_soft_s} and \cref{eq:C_Nash_soft_s}, and the Quantum Alternating Operator Ansatz solutions to \cref{eq:C_social_hard_s} and \cref{eq:C_Nash_hard_s}.

\cref{tab:ng_correct_out_of_ten} shows the number of experimental runs in which the most likely state was also the optimal solution. Several trends are evident in this table. First, increasing $p$ tends to improve performance, consistent with the theoretical expectation of QAOA. Second, the Quantum Alternating Operator Ansatz out-performed its classic counterpart, which we attribute to the reduction in search space enacted by the parity mixers. Finally, the optimal social solution is easier to locate for Quantum Alternating Operator Ansatz implementation than the optimal Nash equilibrium. For the classic variant of QAOA, the reverse is true.

\begin{table}[h]
	\caption{Number of simulations where the system state with the highest probability of measurement was the optimal for the asymmetric network game. 10 simulation runs were performed in total.}
	\label{tab:ng_correct_out_of_ten}
	\centering
    \begin{tabular}{ccccc}
		\toprule
		Quantum & \multicolumn{2}{c}{Optimal} & \multicolumn{2}{c}{Optimal} \\
		Steps & \multicolumn{2}{c}{Social Solution} & \multicolumn{2}{c}{Nash Equilibrium} \\
		($p$) & \emph{classic} & \emph{ansatz} & \emph{classic} & \emph{ansatz} \\
		\midrule
		1 & 0 & 3 & 1 & 7 \\
		2 & 0 & 6 & 1 & 0 \\
		3 & 1 & 8 & 1 & 1 \\
		4 & 1 & 10 & 1 & 7 \\
		5 & 0 & 10 & 2 & 9 \\
		8 & 3 & 10 & 3 & 10 \\
		\bottomrule
    \end{tabular}
\end{table}

\begin{figure}
	\centering
	\includegraphics[width=\linewidth]{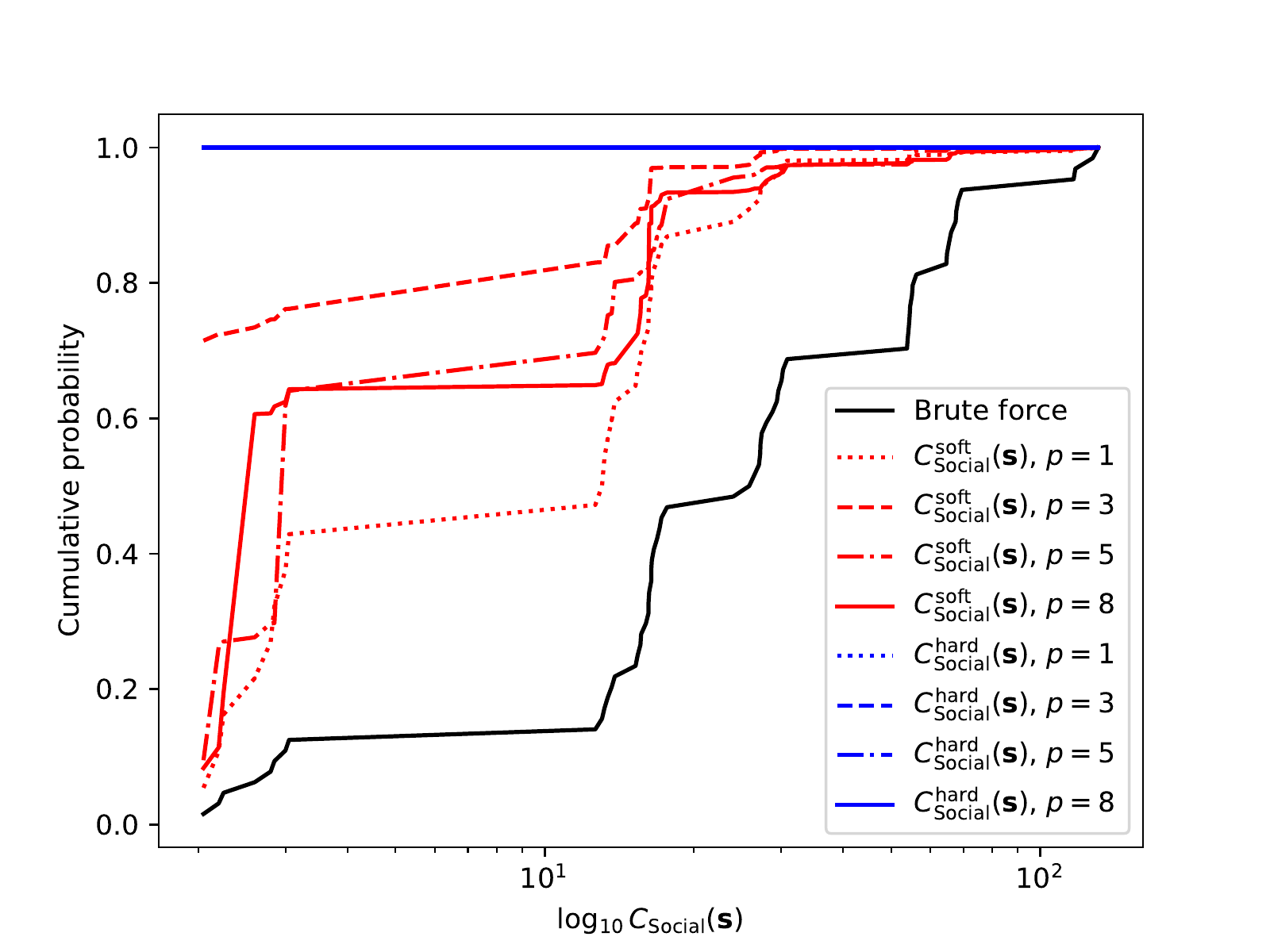}
	\caption{Cumulative probability of measurement for the asymmetric network game for both QAOA variants and varying parameterized repetitions $p$, for all \textbf{possible} solutions to the \textbf{optimal social solution} against brute force results.}
	\label{fig:social_all_possible_solutions}

	\centering
	\includegraphics[width=\linewidth]{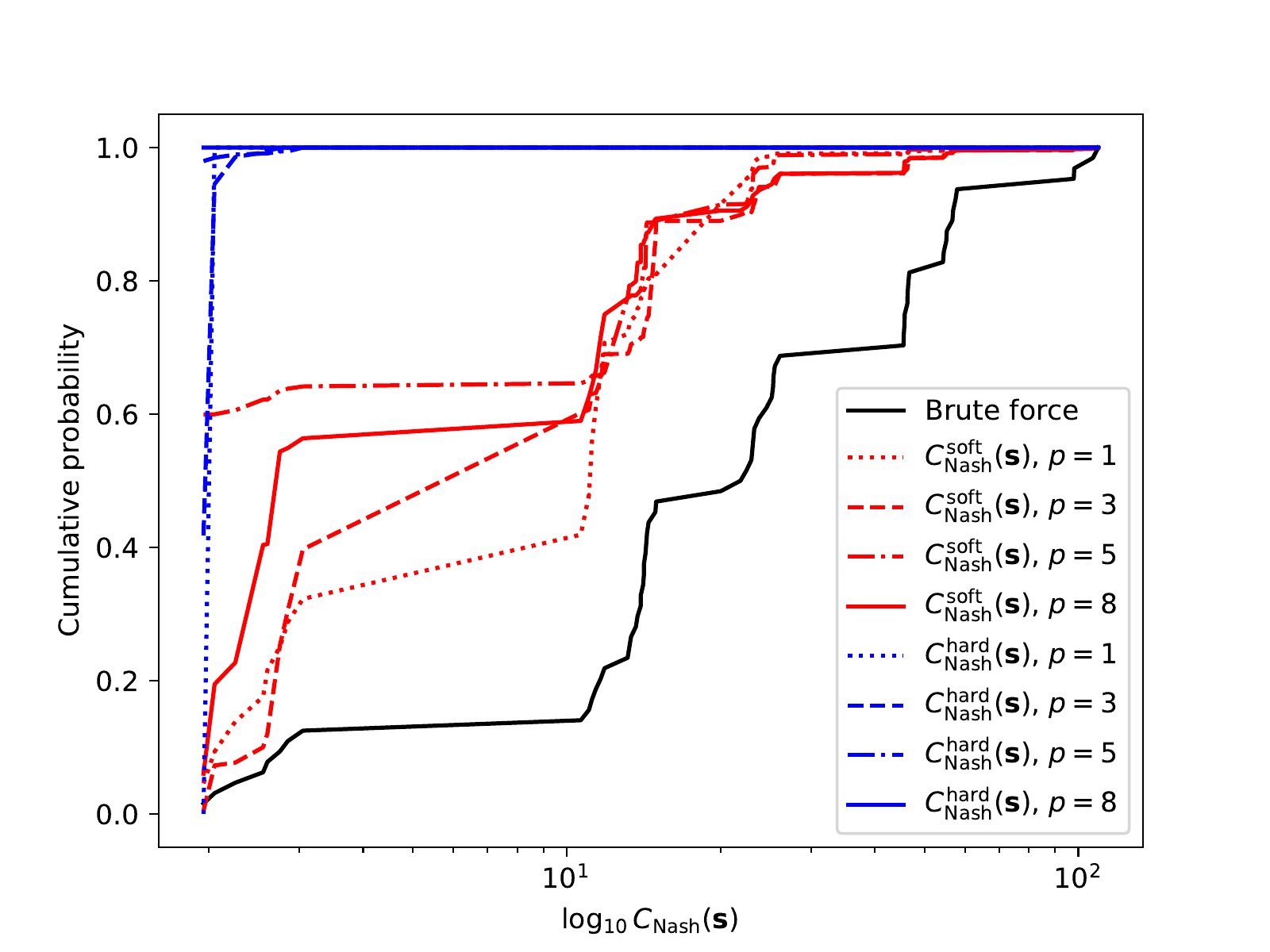}
	\caption{Cumulative probability of measurement for the asymmetric network game for both QAOA variants and varying parameterized repetitions $p$, for all \textbf{possible} solutions to the \textbf{optimal Nash equilibrium} against brute force results.}
	\label{fig:nash_all_possible_solutions}
\end{figure}

\cref{fig:social_all_possible_solutions} and \cref{fig:nash_all_possible_solutions} show the cumulative probability of measurement in the space of all 64 solutions, ordered by cost function. Both QAOA variants show a significant improvement in results compared to random draw from the solution space, represented by the brute force distribution. As expected, there is a performance advantage observed for the hard constraint formulation. In our specific experimental run, the hard constraint optimization found the optimal social solution for all values of $p$. We do not expect this result to scale, however, as the feasible solution space is very small, and the initial state is a mixed state created from the computational basis state $|1\rangle$ which happens to be the social optimal solution, and for selected values of $\beta$ may have biased the outcome.

\begin{figure}
	\centering
	\includegraphics[width=\linewidth]{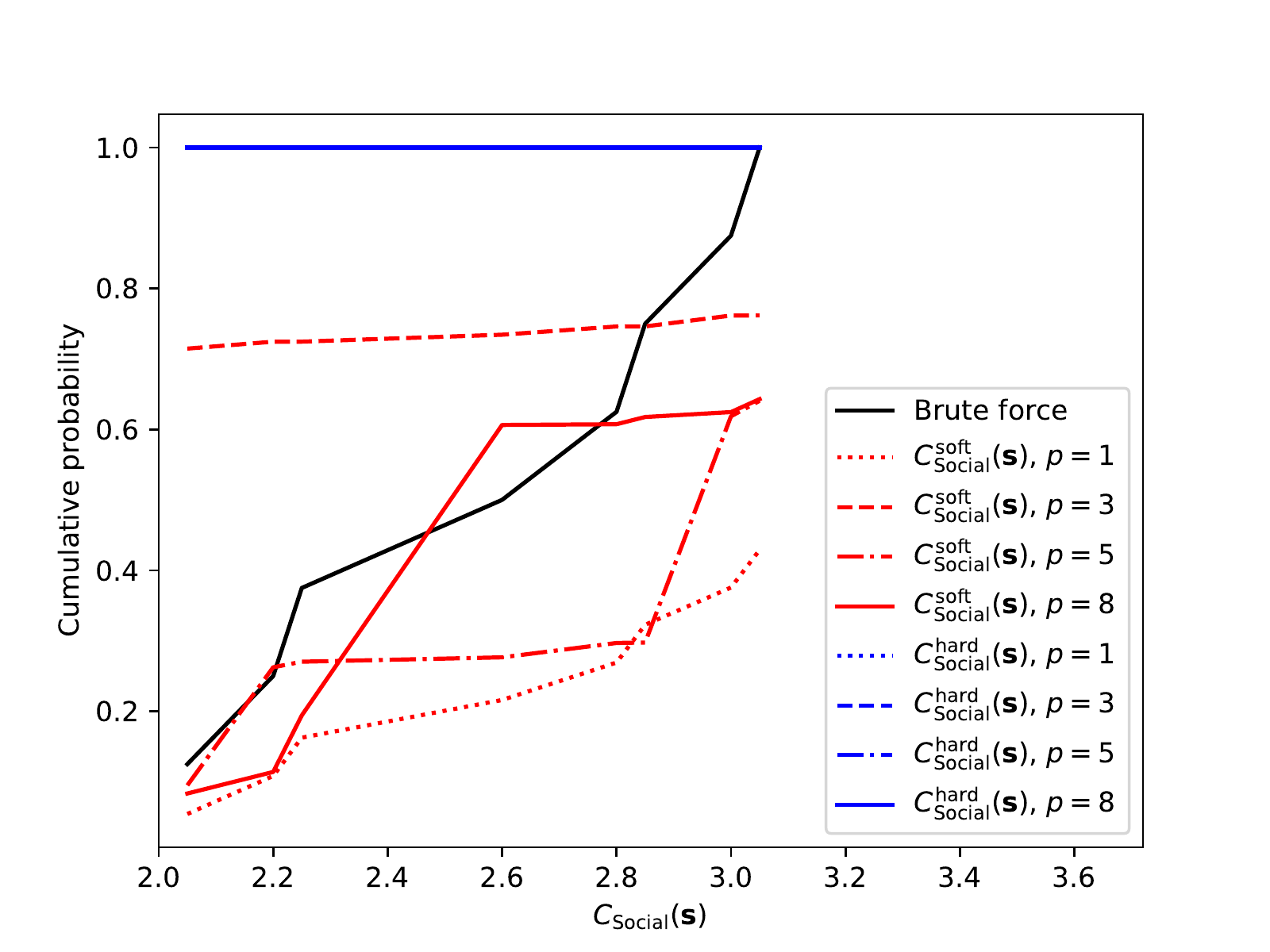}
	\caption{Cumulative probability of measurement for the asymmetric network game for both QAOA variants and varying parameterized repetitions $p$, for all \textbf{feasible} solutions to the \textbf{optimal social solution} against brute force results.}
	\label{fig:social_all_feasible_solutions}

	\centering
	\includegraphics[width=\linewidth]{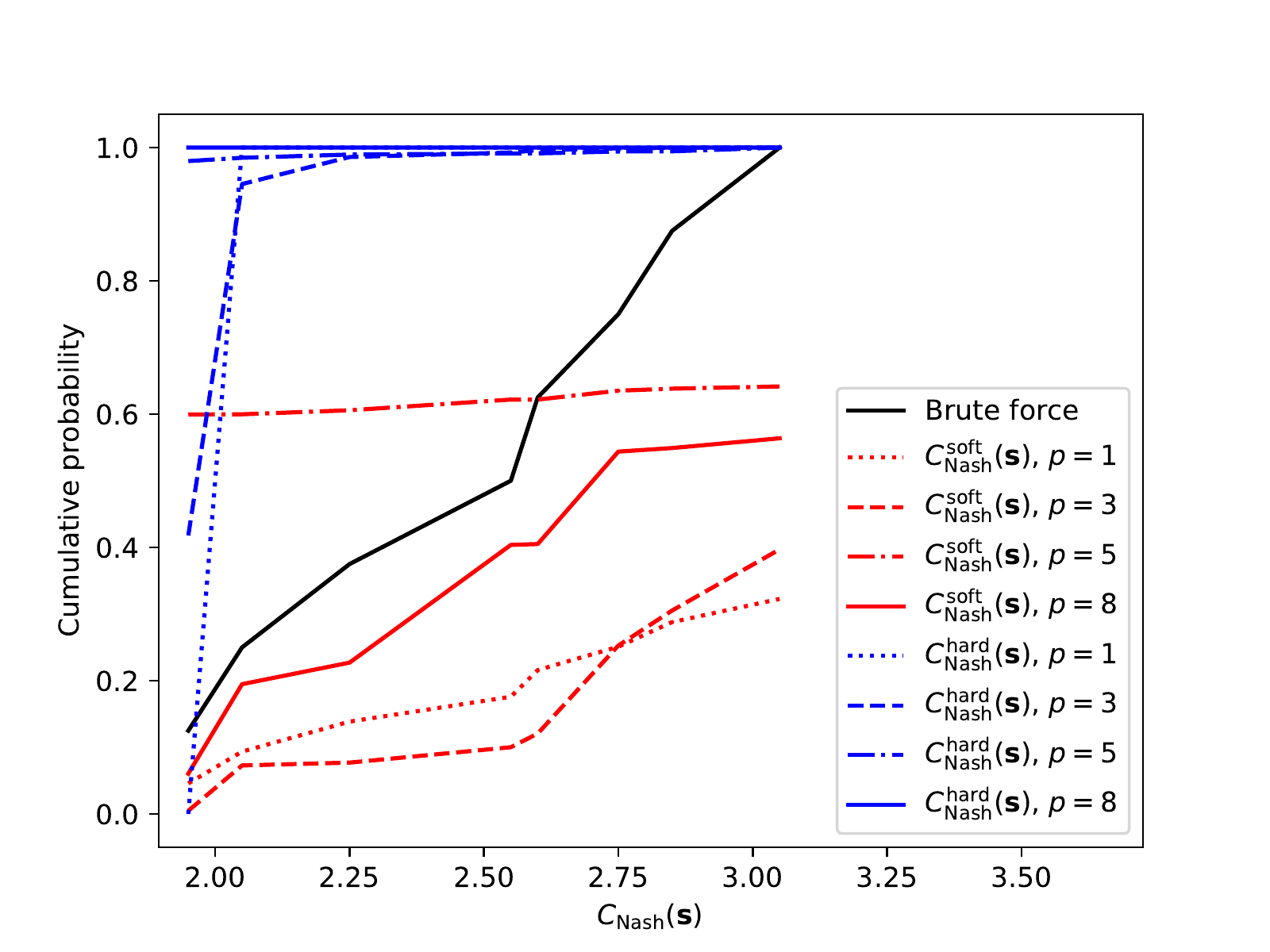}
	\caption{Cumulative probability of measurement for the asymmetric network game for both QAOA variants and varying parameterized repetitions $p$, for all \textbf{feasible} solutions to the \textbf{optimal Nash equilibrium} against brute force results.}
	\label{fig:nash_all_feasible_solutions}
\end{figure}

\cref{fig:social_all_feasible_solutions} and \cref{fig:nash_all_feasible_solutions} show the cumulative probability of measurement in the space of the 8 feasible solutions, ordered by cost function. These plots allow us to observe that the Quantum Alternating Operator Ansatz out-performs random draw even limited to the feasible solution space. Using the hard constraint formulation, the optimal Nash equilibrium was observed with 0\% probability at $p=1$, increasing to 100\% at $p=8$. Using the soft constraint formulation, the optimal result was achieved between 0.5\% and 60\% probability, with the outlier at $p=5$ attributed to the sensitivity of the initial parameter setting and the use of only 10 random seeds.

\section{Conclusion and Future Work}
\label{sec:machine_learning_conclusion}

In this work we have shown the application of QAOA to an asymmetric congestion game, solving for both the optimal social solution and the optimal Nash equilibrium. We used the potential function approach to solving the optimal Nash equilibrium, which is an approach shared by recent research into solving generative adversarial networks, and may open new pathways to quantum assisted machine learning. We prepared a soft constraint formulation based on the Quantum Approximate Optimization Algorithm, and a hard constraint formulation based on the Quantum Alternating Operator Ansatz. We undertook an initial experimental campaign on an idealized simulator of a gate-model quantum computer to verify our implementation and to establish tractability of the approach.

The experimental results are not of sufficient scale to draw conclusions regarding the performance of QAOA in solving an asymmetric congestion game. They do however demonstrate the tractability of the problem to be solved using the potential function approach. This provides a valuable framework for future work in the space of discrete games and generative adversarial networks, where this applied research can be combined with previous works cited in \cref{sec:congestion_game_application_previous_work} to address challenges of training generative adversarial networks involving discrete variables and multi-model distributions.

Lessons learned during experimentation included the importance of pre-mixing in the Quantum Alternating Operator Ansatz when parity mixers are initialized into computational basis states. We also uncovered the importance of randomization of initial state in the parity mixers, which was observed in the analysis stage to have potentially biased our results, creating a perfect 100\% probability of the optimal social solution in some cases.

We recommend two avenues of further investigation. The first is to deepen the development of the congestion game created for this initial investigation. This could include tailoring the formulation to an industrial use case, and analyzing the computational complexity both theoretically and experimentally with respect to a classical benchmark. Experimentation should proceed through idealized simulated resources, simulated noise models and resource estimators, and validation on current-generation NISQ hardware. The second avenue of investigation is to apply the results of this work to training of generative adversarial networks using potential function approaches, extending current research in quantum assisted machine learning to address industrial classification and synthetic data generation problems.


\section{Appendix: Analysis of Limiting Behavior of Quadratic Terms in the Cost Functions}
\label{sec:machine_learning_appendix_big_o}

We provide the details of our analysis of the limiting behavior of the three principal QAOA spin-system cost functions derived in this paper. In this analysis we simplify the spin variable notation from $s_{i,j}$ to $s$, and eliminate all constant factors and lower-order terms in the equations.

\subsection{Analysis of the optimal social solution}
\label{sec:analysis_of_the_optimal_social_solution}

We observe in \cref{eq:congestion_game_number_of_players_s} and \cref{eq:congestion_game_resource_delay_s} that $n_k(\mathbf{s})$ and $d_k(n_k(\mathbf{s}))$ have the same form in the limit once constant factors are removed. This allows us to expand and simplify cost function $C_{\text{social}}(\mathbf{s})$ of \cref{eq:congestion_game_C_s_social} as

\begin{equation*}
	\begin{alignedat}{2}
		C_{\text{social}}(\mathbf{s}) \, & = \sum_{k \in \mathcal{R}} n_k(\mathbf{s}) \, d_k(n_k(\mathbf{s})) \\
		& \sim \sum_{k \in \mathcal{R}} \left( \sum_{i \in \mathcal{N}} \sum_{j \in \mathcal{S}_i \mid k \in j} s \right)^2 \\
	\end{alignedat}
\end{equation*}

Inside the squared term, the number of linear terms that may result is $\mathcal{O}(np)$ where $p \sim |\mathcal{S}_i|$ is the number of paths available to each player. Therefore by inspection the behavior in the limit of the number of quadratic terms that may result in $C_{\text{social}}(\mathbf{s})$ is $\mathcal{O}(r n^2 p^2)$.

\subsection{Analysis of the optimal Nash equilibrium}
\label{sec:analysis_of_the_optimal_nash_equilibrium}

We observe in \cref{eq:congestion_game_C_s_Nash} that the inner summation of the cost function for finding the optimal Nash equilibrium is of a form that can be arranged to take advantage of the triangular number identity

\begin{equation*}
	T_n = \sum_{k=1}^n k = 1 + 2 + 3 + \cdots + n = \frac{n(n+1)}{2}
\end{equation*}

through the steps

\begin{equation*}
	\begin{alignedat}{2}
		C_{\text{Nash}}(\mathbf{s}) \, & = \sum_{k \in \mathcal{R}} \sum_{j}^{n_k(\mathbf{s})} d_k(j) \\
		& = \sum_{k \in \mathcal{R}} \sum_{j}^{n_k(\mathbf{s})} (a_k + b_k j) \\
		& = \sum_{k \in \mathcal{R}} \left( a_k n_k(\mathbf{s}) + b_k \sum_{j}^{n_k(\mathbf{s})} j \right) \\
		& = \sum_{k \in \mathcal{R}} \left( a_k n_k(\mathbf{s}) + b_k T_{n_k(\mathbf{s})} \right) \\
	\end{alignedat}
\end{equation*}

This allows us to expand and simplify cost function $C_{\text{Nash}}(\mathbf{s})$ of \cref{eq:congestion_game_C_s_Nash} into a form that can be analyzed in the limit, as

\begin{equation*}
	\begin{alignedat}{2}
		C_{\text{Nash}}(\mathbf{s}) \, & = \sum_{k \in \mathcal{R}} \left( a_k n_k(\mathbf{s}) + b_k \frac{n_k(\mathbf{s}) (n_k(\mathbf{s}) + 1)}{2} \right) \\
		& \sim \sum_{k \in \mathcal{R}} \left( \sum_{i \in \mathcal{N}} \sum_{j \in \mathcal{S}_i \mid k \in j} s \right)^2 \\
	\end{alignedat}
\end{equation*}

This presents us with the same behavior in the limit as the optimal social solution of \cref{sec:analysis_of_the_optimal_social_solution}. Therefore the behavior in the limit of the number of quadratic terms that may result in $C_{\text{Nash}}(\mathbf{s})$ is $\mathcal{O}(r n^2 p^2)$.

\subsection{Analysis of the soft constraint penalty function}
\label{sec:analysis_of_the_soft_constraint_penalty_function}

We analyze the behavior in the limit for soft constraint penalty function designed to enforce the path selection constraint, \cref{eq:congestion_game_path_constraint_s}, as

\begin{equation*}
	\begin{alignedat}{2}
		C_{\text{path}}(\mathbf{s}) \, & = A \sum_{i \in \mathcal{N}} \left( \sum_{j \in S_i} s_{i,j} + |\mathcal{S}_i| - 2 \right)^2 \\
		& \sim \sum_{i \in \mathcal{N}} \left( \sum_{j \in S_i} s \right)^2 \\
	\end{alignedat}
\end{equation*}

Therefore the behavior in the limit of the number of quadratic terms that may result in $C_{\text{path}}(\mathbf{s})$ is $\mathcal{O}(n p^2)$.

\subsection{A note on analysis in the limit}

The analysis in the limit for the two optimization functions assumes a worst case. It does not account for the possibility of interplay between the actual number of player paths sharing a resource (the sum over $j \in \mathcal{S}_i \mid k \in j$), and the total number of resources in the outer sums (the sum over $k \in \mathcal{R}$). For specific networks with additional topological assumptions, this worst case may be able to be reduced.

\renewcommand*{\UrlFont}{\rmfamily}  
\printbibliography

@ARTICLE{Hui2018,
       author = {Hui, Jonathan},
        title = "{GAN - Why it is so hard to train Generative Adversarial Networks!}",
      journal = "Medium",
         year = "2018",
        month = "6",
          URL = { https://medium.com/@jonathan_hui/gan-why-it-is-so-hard-to-train-generative-advisory-networks-819a86b3750b }
}

@ARTICLE{Kondratyev2019,
       author = {Kondratyev, Alexei and Schwarz, Christian},
        title = "{The Market Generator}",
      journal = {SSRN},
         year = "2019",
        month = "5",
          doi = {10.2139/ssrn.3384948},
          URL = { http://dx.doi.org/10.1080/10556780701722542 }
}

@ARTICLE{Zenati2018,
       author = {{Zenati}, Houssam and {Foo}, Chuan Sheng and {Lecouat}, Bruno and
         {Manek}, Gaurav and {Ramaseshan Chandrasekhar}, Vijay},
        title = "{Efficient GAN-Based Anomaly Detection}",
      journal = {arXiv e-prints},
         year = "2018",
        month = "2",
archivePrefix = {arXiv},
       eprint = {1802.06222},
 primaryClass = {cs.LG},
       adsurl = {https://ui.adsabs.harvard.edu/abs/2018arXiv180206222Z}
}

@ARTICLE{Shah2018,
       author = {Shah, Hamaad},
        title = "{Using Bidirectional Generative Adversarial Networks to estimate Value-at-Risk for Market Risk Management}",
      journal = "Towards Data Science",
         year = "2018",
        month = "8",
          URL = { https://towardsdatascience.com/using-bidirectional-generative-adversarial-networks-to-estimate-value-at-risk-for-market-risk-c3dffbbde8dd }
}

@ARTICLE{Koshiyama2019,
       author = {{Koshiyama}, Adriano and {Firoozye}, Nick and {Treleaven}, Philip},
        title = "{Generative Adversarial Networks for Financial Trading Strategies Fine-Tuning and Combination}",
      journal = {arXiv e-prints},
         year = "2019",
        month = "1",
archivePrefix = {arXiv},
       eprint = {1901.01751},
 primaryClass = {cs.LG},
       adsurl = {https://ui.adsabs.harvard.edu/abs/2019arXiv190101751K}
}

@ARTICLE{Hadad2017,
       author = {{Hadad}, Naama and {Wolf}, Lior and {Shahar}, Moni},
        title = "{Two-Step Disentanglement for Financial Data}",
      journal = {arXiv e-prints},
         year = "2017",
        month = "9",
archivePrefix = {arXiv},
       eprint = {1709.00199},
 primaryClass = {cs.LG},
       adsurl = {https://ui.adsabs.harvard.edu/abs/2017arXiv170900199H}
}

@ARTICLE{DelPia2016,
       author = {{Del Pia}, Alberto and {Ferris}, Michael and {Michini}, Carla},
        title = "{Totally Unimodular Congestion Games}",
      journal = {arXiv e-prints},
         year = "2015",
        month = "11",
archivePrefix = {arXiv},
       eprint = {1511.02784},
 primaryClass = {cs.GT},
       adsurl = {https://ui.adsabs.harvard.edu/abs/2015arXiv151102784D}
}

@ARTICLE{Mahabubul2019,
       author = {{Alam}, Mahabubul and {Ash-Saki}, Abdullah and {Ghosh}, Swaroop},
        title = "{Analysis of Quantum Approximate Optimization Algorithm under Realistic Noise in Superconducting Qubits}",
      journal = {arXiv e-prints},
         year = "2019",
        month = "7",
archivePrefix = {arXiv},
       eprint = {1907.09631},
 primaryClass = {quant-ph},
       adsurl = {https://ui.adsabs.harvard.edu/abs/2019arXiv190709631A}
}

@article{Smith2017,
	title={{CBA} steps into the future with quantum computing simulator},
	url={https://www.afr.com/technology/cba-steps-into-the-future-with-quantum-computing-simulator-20170407-gvg52l},
	journal={The Australian Financial Review},
	author={Smith, Paul},
	year={2017},
	month={4}
}

@ARTICLE{Preskill2018,
       author = {{Preskill}, John},
        title = "{Quantum Computing in the NISQ era and beyond}",
      journal = {arXiv e-prints},
         year = {2018},
        month = {1},
archivePrefix = {arXiv},
       eprint = {1801.00862},
 primaryClass = {quant-ph},
       adsurl = {https://ui.adsabs.harvard.edu/abs/2018arXiv180100862P}
}

@article{Farhi2014,
       author = "Farhi, E. and Goldstone, J. and Gutmann, S.",
        title = "A Quantum Approximate Optimization Algorithm",
archivePrefix = "arXiv",
       eprint = "1411.4028",
 primaryClass = "quant-ph",
         year = "2014",
		  url = "https://arxiv.org/abs/1411.4028"
}

@article{Wang2017,
       author = "Wang, Z. and Hadfield, S. and Jiang, Z. and Rieffel, E.G.",
        title = "The Quantum Approximization Algorithm for MaxCut: A Fermionic View",
archivePrefix = "arXiv",
       eprint = "1706.02998",
 primaryClass = "quant-ph",
         year = "2017",
		  url = "https://arxiv.org/abs/1706.02998"
}

@Article{Rosenthal1973,
   author = "Rosenthal, Robert W.",
    title = "A class of games possessing pure-strategy {Nash} equilibria",
  journal = "International Journal of Game Theory",
     year = {1973},
    month = {12},
      day = 1,
   volume = 2,
   number = 1,
    pages = "65--67",
     issn = "1432-1270",
      doi = "10.1007/BF01737559",
      url = "https://doi.org/10.1007/BF01737559"
}

@article{YANNAKAKIS200971,
title = "Equilibria, fixed points, and complexity classes",
journal = "Computer Science Review",
volume = "3",
number = "2",
pages = "71 - 85",
year = "2009",
issn = "1574-0137",
doi = "https://doi.org/10.1016/j.cosrev.2009.03.004",
url = "http://www.sciencedirect.com/science/article/pii/S1574013709000161",
author = "Mihalis Yannakakis"
}

@inproceedings{Conitzer:2003:CRN:1630659.1630770,
 author = {Conitzer, Vincent and Sandholm, T\"{u}omas},
 title = {Complexity Results About {Nash} Equilibria},
 booktitle = {Proceedings of the 18th International Joint Conference on Artificial Intelligence},
 series = {IJCAI'03},
 year = {2003},
 location = {Acapulco, Mexico},
 pages = {765--771},
 numpages = {7},
 url = {http://dl.acm.org/citation.cfm?id=1630659.1630770},
 acmid = {1630770},
 publisher = {Morgan Kaufmann Publishers Inc.},
 address = {San Francisco, CA, USA},
}

@inproceedings{Conitzer:2006:COS:1134707.1134717,
 author = {Conitzer, Vincent and Sandholm, Tuomas},
 title = {Computing the Optimal Strategy to Commit to},
 booktitle = {Proceedings of the 7th ACM Conference on Electronic Commerce},
 series = {EC '06},
 year = {2006},
 isbn = {1-59593-236-4},
 location = {Ann Arbor, Michigan, USA},
 pages = {82--90},
 numpages = {9},
 url = {http://doi.acm.org/10.1145/1134707.1134717},
 doi = {10.1145/1134707.1134717},
 acmid = {1134717},
 publisher = {ACM},
 address = {New York, NY, USA},
}

@misc{Dutting2015,
 author = {D{\"u}tting, P.},
 title = {Complexity of Pure {Nash} Equilibria in Congestion Games},
 bookTitle = "{Algorithmic Game Theory, Summer Week 2}",
 year = {2015},
 publisher = {ETH Z{\"u}rich},
 url = {https://www.cadmo.ethz.ch/education/lectures/HS15/agt_HS2015/agt_hs2015_lec02_complexity.pdf}
}

@techreport{Meyers2008,
 author = {Myers, C.A. and Schulz, A.S.},
 title = {The complexity of congestion games},
 institution = {Massachusetts Institute of Technology, Cambridge},
 year = {2008},
 url = {http://web.mit.edu/schulz/www/epapers/meyers-schulz-june-2008.pdf},
}

@article{MondererShapley1996,
title = "Potential Games",
journal = "Games and Economic Behavior",
volume = "14",
number = "1",
pages = "124-143",
year = "1996",
doi = "http://dx.doi.org/10.1006/game.1996.0044",
author = "Monderer, D. and Shapley, L.S."
}

@article{PerdomoOrtizEtAl2018,
author={Perdomo-Ortiz, A. and Benedetti, M. and Realpe-G\'{o}mez, J. and Biswas, R.},
title={Opportunities and challenges for quantum-assisted machine learning in near-term quantum
computers},
journal={Quantum Sci. Technol.},
volume={3},
number={3},
pages={},
url={http://stacks.iop.org/2058-9565/3/i=3/a=030502},
year={2018}
}

@article{GoodfellowEtAl2014,
author={Ian Goodfellow and Jean Pouget-Abadie and Mehdi Mirza and Bing Xu and David Warde-Farley and Sherjil Ozair and Aaron Courville and Yoshua Bengio},
title={Generative adversarial nets},
journal={Advances in neural information processing systems},
volume={3},
number={3},
pages={2672-2680},
url={ https://arxiv.org/pdf/1406.2661.pdf},
year={2014},
}

@article{BenedettiEtAl2018,
       author = "Marcello Benedetti and Edward Grant and Leonard Wossnig and Simone Severini",
        title = "Adversarial quantum circuit learning for pure state approximation",
archivePrefix = "arXiv",
       eprint = "1806.00463",
 primaryClass = "quant-ph",
         year = "2018",
		  url = "https://arxiv.org/abs/1806.00463"
}

@article{DallaireDemersKilloran2018,
       author = "Pierre-Luc Dallaire-Demers and Nathan Killoran",
        title = "Quantum generative adversarial networks",
archivePrefix = "arXiv",
       eprint = "1804.08641",
 primaryClass = "quant-ph",
         year = "2018",
		  url = "https://arxiv.org/abs/1804.08641"
}

@article{LloydWeedbrook2018,
       author = "Seth Lloyd and Christian Weedbrook",
        title = "Quantum generative adversarial learning",
archivePrefix = "arXiv",
       eprint = "1804.09139",
 primaryClass = "quant-ph",
         year = "2018",
		  url = "https://arxiv.org/abs/1804.09139"
}

@article{OliehoekEtAl2018,
       author = "Oliehoek, F.A. and Savani, R. and Gallego, J. and van der Pol, E. and Gro{\ss}, R.",
        title = "Beyond Local {Nash} Equlibria for Adversarial Networks",
archivePrefix = "arXiv",
       eprint = "1806.07268",
 primaryClass = "cs.GT",
         year = "2018",
		  url = "https://arxiv.org/pdf/1806.07268.pdf"
}

@article{Hadfield2019,
	title = {From the Quantum Approximate Optimization Algorithm to a Quantum Alternating Operator Ansatz},
	volume = {12},
	rights = {http://creativecommons.org/licenses/by/3.0/},
	url = {https://www.mdpi.com/1999-4893/12/2/34},
	doi = {10.3390/a12020034},
	pages = {34},
	number = {2},
	journaltitle = {Algorithms},
	author = {Hadfield, Stuart and Wang, Zhihui and O'Gorman, Bryan and Rieffel, Eleanor G. and Venturelli, Davide and Biswas, Rupak},
	urldate = {2019-04-10},
	date = {2019-02},
	langid = {english}
}

@ARTICLE{Lucas2014,
       author = {{Lucas}, Andrew},
        title = "{Ising formulations of many NP problems}",
      journal = {Frontiers in Physics},
         year = {2014},
        month = {2},
       volume = {2},
          eid = {5},
        pages = {5},
          doi = {10.3389/fphy.2014.00005},
archivePrefix = {arXiv},
       eprint = {1302.5843},
 primaryClass = {cond-mat.stat-mech},
       adsurl = {https://ui.adsabs.harvard.edu/abs/2014FrP.....2....5L}
}

\end{document}